\documentclass[a4paper,fleqn,usenatbib]{mnras}

\usepackage{mathptmx}

\usepackage[T1]{fontenc}
\usepackage{ae,aecompl}
\usepackage{tablefootnote}

\usepackage{graphicx}	
\usepackage[fleqn]{amsmath}	
\usepackage{amssymb}	
\usepackage{subfig}

\title[RCW$\,$38]{The low-mass content of the massive young star cluster RCW\,38\thanks{Based on observations collected at the European Southern 
Observatory under programmes 090.C-270 and 70.C-0400}}

\author[K. Mu\v{z}i\'c et al.]{
Koraljka Mu\v{z}i\'c$^{1,2}$\thanks{E-mail: kmuzic@sim.ul.pt},
Rainer Sch\"{o}del$^{3}$,
Alexander Scholz$^{4}$,
Vincent C. Geers$^{5}$,
\newauthor
Ray Jayawardhana$^{6}$,
Joana Ascenso$^{7,8}$
and Lucas A. Cieza$^{1}$
\\
$^{1}$N\'ucleo de Astronom\'ia, Facultad de Ingenier\'ia, Universidad Diego Portales, Av. Ejercito 441, Santiago, Chile\\
$^{2}$SIM/CENTRA, Faculdade de Ciencias de Universidade de Lisboa, Ed. C8, Campo Grande, P-1749-016 Lisboa, Portugal\\
$^{3}$Instituto de Astrof\'isica de Andaluc\'ia (CSIC), Glorieta de la Astronom\'a s/n, 18008 Granada, Spain\\
$^{4}$SUPA, School of Physics \& Astronomy, St. Andrews University, North Haugh, St Andrews KY16 9SS, United Kingdom\\
$^{5}$UK Astronomy Technology Centre, Royal Observatory Edinburgh, Blackford Hill, Edinburgh, EH9 3HJ, United Kingdom\\
$^{6}$Faculty of Science, York University, 355 Lumbers Building, 4700 Keele Street, Toronto, ON M3J 1P2, Canada\\
$^{7}$CENTRA, Instituto Superior Tecnico, Universidade de Lisboa, Av. Rovisco Pais 1, 1049-001 Lisbon, Portugal\\
$^{8}$Universidade do Porto, Departamento de Engenharia F\'isica da Faculdade de Engenharia, Rua Dr. Roberto Frias, s/n, \\P-4200-465, Porto, Portugal
}
\date{Accepted XXX. Received YYY; in original form ZZZ}

\pubyear{2017}

\begin{document}
\label{firstpage}
\pagerange{\pageref{firstpage}--\pageref{lastpage}}
\maketitle

\begin{abstract}
RCW\,38 is a deeply embedded young ($\sim$1 Myr), massive star cluster located at a distance of 1.7 kpc. 
Twice as dense as the Orion Nebula Cluster, orders of magnitude denser than other nearby star forming regions, and rich in massive stars, 
RCW\,38 is an ideal place to look for potential differences in brown dwarf formation efficiency as a function of environment.
We present deep, high resolution adaptive optics data of the central $\sim0.5\times0.5$\,pc$^2$ obtained with NACO at the Very Large Telescope.
Through comparison with evolutionary models we determine masses and extinction for $\sim 480$ candidate members, and 
derive the first Initial Mass Function (IMF) of the cluster extending into the substellar regime.
Representing the IMF as a set of power laws in the form $dN/dM \propto M^{-\alpha}$, we derive
the slope $\alpha = 1.60 \pm 0.13$ for the mass range 0.5 -- 20\,M$_{\sun}$
which is shallower than the Salpeter slope, but 
in agreement with results in several other young massive clusters. At the low-mass side, we find $\alpha = 0.71 \pm 0.11$ for masses between 0.02 and 0.5\,M$_{\sun}$,
or $\alpha = 0.81 \pm 0.08$ for masses between 0.02 and 1\,M$_{\sun}$. Our result is in agreement with the values found in other young star-forming regions, 
revealing no evidence that a combination of high stellar densities and the presence of numerous massive stars 
affect the formation efficiency of brown dwarfs and very-low mass stars. 
 We estimate that the Milky Way galaxy contains between 25 and 100 billion brown dwarfs (with masses > 0.03\,M$_{\sun}$).

\end{abstract}




\section{Introduction}

Stellar clusters born embedded in molecular clouds have long been recognized as important laboratories for understanding how star formation works.
They give birth to stars across a wide range of masses, from the most massive stars reaching several tens of 
solar masses, to the substellar objects with masses below 0.075\,M$_{\sun}$. The distribution of 
masses in young clusters, the initial mass function (IMF), is a fundamental outcome of star formation, and therefore has 
been a topic of extensive research. 
Measurements of the IMF have been represented in the literature by various functional forms.
On the high mass side, the mass function can be described as a 
power law $dN/dM \propto M^{-\alpha}$
with the Salpeter slope $\alpha = 2.35$ \citep{salpeter55}. 
As the studies of the Solar neighbourhood and the Galactic disk show, the Salpeter slope 
seems to be a good approximation of the mass function down to $\sim\,$0.5\,M$_{\sun}$, but below this mass
the distribution becomes significantly flatter (by $\Delta \alpha \sim\,$1.5; \citealt{luhman12} and references therein). 
This lead to updated descriptions of the IMF, such as a log-normal distribution \citep{chabrier05}, 
a set of power-law functions \citep{kroupa01}, or a tapered power-law \citep{demarchi05}.
The flattening of the IMF at the low masses is also seen in young clusters and star forming regions, which 
in general yield slopes $\alpha < 1$,
for the masses below $\sim 1$\,M$_{\sun}$ (e.g., \citealt{luhman04c, luhman07, bayo11, penaramirez12, scholz12a, lodieu13, muzic15}).

The low-mass end of the IMF is the focus of  
the SONYC survey - short for {\it Substellar Objects in Nearby Young Clusters} - our
project aiming to provide a complete, unbiased census of the substellar population
in nearby star forming regions. 
The survey is based on deep imaging on 4-8-m-class telescopes, and followed up with extensive spectroscopic campaigns targeting candidate very-low mass objects
in four nearby star forming regions. We reach completeness limits of 0.005 - 0.01\,M$_{\sun}$ in NGC\,1333 and Cha-I, 
0.1 - 0.2\,M$_{\sun}$ in Lupus 3, and $\sim$0.03\,M$_{\sun}$ in $\rho$-Oph. 
Thanks to SONYC, and similar recent surveys by other groups, the substellar IMF in nearby regions is now well characterized 
down to $\sim$10\,M$_{Jup}$, and in a few selected
regions even deeper ($\sim$5\,M$_{Jup}$). In SONYC, we find that for every formed brown dwarf (BD), there are 2 - 6 stars, and the slope
of the mass function below $1$\,M$_{\sun}$ is $\alpha\sim0.7$ \citep{scholz12b, scholz13, muzic15}. The relatively large range in the star to BD number ratio 
does not necessarily reflect differences between different regions, it is rather caused by 
uncertainties due to incompleteness of the spectroscopic follow-up, choice of isochrones
used to derive masses, uncertainties in distances, etc.
The IMF in most star forming regions appears to be continuous across the hydrogen- and deuterium-burning limits.
In a comparative study of seven nearby  star forming regions, \citet{andersen08} find that the ratios of stars
to brown dwarfs are consistent with a single underlying IMF. 

While strong IMF variations are excluded for most nearby regions, there might a few notable exceptions, that hint 
 that the environment might after all influence the BD production efficiency.
In \citet{scholz13}, we investigated 
the mass functions in IC\,348 and NGC\,1333, two young clusters in Perseus. We find that, under plausible assumptions, the cumulative mass distributions 
of the two clusters are significantly different, resulting from an overabundance of very-low-mass objects in NGC\,1333. 
In a recent analysis of the same two clusters, \citet{luhman16} also report a higher abundance of low-mass objects in NGC\,1333.
This  suggests that the relative number of very-low-mass objects in star forming regions might 
depend on stellar density, in the sense that regions with higher densities produce more objects with very low masses.  Furthermore, in Lupus 3, we find evidence 
for a turn-down of the IMF in the substellar regime. This cluster seems to produce fewer BDs than other 
clusters we have studied \citep{muzic15, comeron11}. 

Most of the star forming regions where the substellar content was studied in detail are relatively loose groups of low
mass stars, with no, or very few massive stars, and mostly similar in their properties. 
An exception is the Orion Nebula Cluster (ONC), the only massive star forming region within 500\,pc distance.
The ONC low-mass IMF has been a subject of several studies to date, yielding somewhat inconsistent results, from
a steeply declining IMF \citep{dario12}, to an upturn in the low-mass regime \citet{muench02}. 
Most recently, based on the new photometric near-infrared observations of the ONC, \citet{drass16} derive a bimodal form of an IMF, 
with a dip around 0.1\,M$_{\sun}$, and two peaks at 0.25 and 0.025\,M$_{\sun}$.
A similar, although shallower dip was previously reported in the analysis by \citet{lucas05}. \citet{drass16} 
interpret the substellar peak as possibly formed by brown dwarfs ejected from multiple systems or circumstellar disks at an early stage of star 
formation.

Environmental difference is in fact theoretically expected, as
most of the current BD formation theories predict an overproduction of substellar objects in dense environments,
or close to very massive stars. \citet{bonnell08} investigate the formation of BDs and very-low-mass stars
through the gravitational fragmentation of infalling gas into stellar clusters, where these clusters provide the gravitational
focus that attracts gas from the surrounding molecular cloud, and provide gas densities that are sufficiently large to form
BDs. BDs in this simulation are therefore preferentially formed in regions of high stellar densities. In the
largest hydrodynamical simulation of cluster formation to date \citep{bate12}, very-low-mass objects are formed by ejection
from the reservoir of accretion material, and the efficiency of their formation is expected to depend on density, as higher
stellar densities favor ejections. Furthermore, the turbulent fragmentation framework also predicts that an enhancement in
density should lead to an increase in the numbers of BDs \citep{padoan02, hennebelle09}.
Finally, a viable channel for BD formation is in the vicinity of OB stars, whose powerful ionization fronts can erode
the outer layers of a pre-stellar core, leaving a small fragment that can only form a substellar object \citep{whitworth04}.

After consolidating the shape of the IMF in most of the nearby star forming regions,
the next logical step to test potential
environmental differences in the production of very-low mass stars and BDs is to analyze their content in some of the massive, dense
embedded clusters.  
The main challenges in studying these clusters are their distances, crowding (in some cases), and high extinctions due to the high column density molecular clouds
with which they are associated. As a result, most studies of the stellar content in massive embedded clusters extend down to
a few solar masses, and only a few reach close to the hydrogen burning limit. The recent study in Westerlund 1 by \citet{andersen16} extends down 
to 0.15\,M$_{\sun}$ in the outer parts of the cluster, an unprecedented depth for such a distant cluster ($\sim 4\,$kpc). 
Similar mass limits have been reached in Trumpler\,14 \citep{rochau11}.
The IMF has been characterized in NGC\,3603 down to 0.4\,M$_{\sun}$ \citep{stolte06, harayama08}.

In this paper, we present new adaptive optics (AO) observations of the central half parsec of the young embedded cluster RCW\,38.
Previous deep observations of the core of this cluster reached down to the hydrogen-burning limit \citep{derose09}, 
and the present data extend this dataset further down into the substellar regime.
RCW\,38 is a young ($<1$\,Myr), embedded cluster surrounding a pair of early O stars, and located
at the distance of 1700\,pc \citep{wolk08}. 
The distance to RCW\,38 has been derived from the photometry of the OB stars candidates \citep{muzzio79, avedisova89}, and
from examining the CO structures in the Vela ridge, in addition to the photometric data \citep{murphy85}. The X-ray luminosity 
function is consistent with the 1700 pc estimate from these works \citep{wolk06}. 
Several works have inferred the very young age of the cluster. \citep{wolk06} estimate the age to be 0.5 Myr, with an upper limit
of 1 Myr (older age doubles the number of O stars, which the authors find unlikely). Using the NIR and X-ray photometry, \citet{getman14}
estimate individual ages of several cluster members, giving on average $\sim\,1$\,Myr. The disk fraction of $\sim 70 \%$ is also consistent
with the same age \citep{winston11}.
The CO maps of the region reveal that the stellar cluster is associated with two molecular clouds containing the total gas mass of 2.3$\times10^4$\,M$_{\sun}$, whose collision possibly
triggered the recent massive star formation \citep{fukui16}.
The massive stellar content and the structure of RCW\,38 have been characterized 
as part of the MYStIX project \citep{kuhn15}, who find that this cluster has the highest projected core surface 
density among all the clusters in their study (20 massive young clusters within 4 kpc),
 and twice as high as that of the ONC. The peak core surface density of RCW\,38 is found to be $\sim\,$34,000 stars pc$^{-2}$\footnote{We will re-derive the surface densities for the cluster core in Section~\ref{densities}.}, compared to 
 $\sim\,$17,000 stars pc$^{-2}$ in the ONC \citep{kuhn15}, 2000\,pc$^{-2}$ in $\rho$-Oph and $\leq$1000 \,pc$^{-2}$ in the remaining star forming regions 
 covered by SONYC \citep{gutermuth09}. The analysis based on the 
X-ray, near-, and mid-infrared data \citep{wolk06, winston11} identified more than 600 YSOs, as well as about 60 O- and OB-star candidates located
in four sub-clusters, and distributed in the area around them. The total extent of the cluster can be approximated by an ellipsoid
with semi-major and -minor axes of four times the cluster core radius, $\sim\,$10\,pc\,$\times$\,8\,pc \citep{kuhn14}. Twelve of the OB-star candidates are found in the central
core of the cluster studied in this work.

\begin{figure*}
\centering
\resizebox{16cm}{!}{\includegraphics{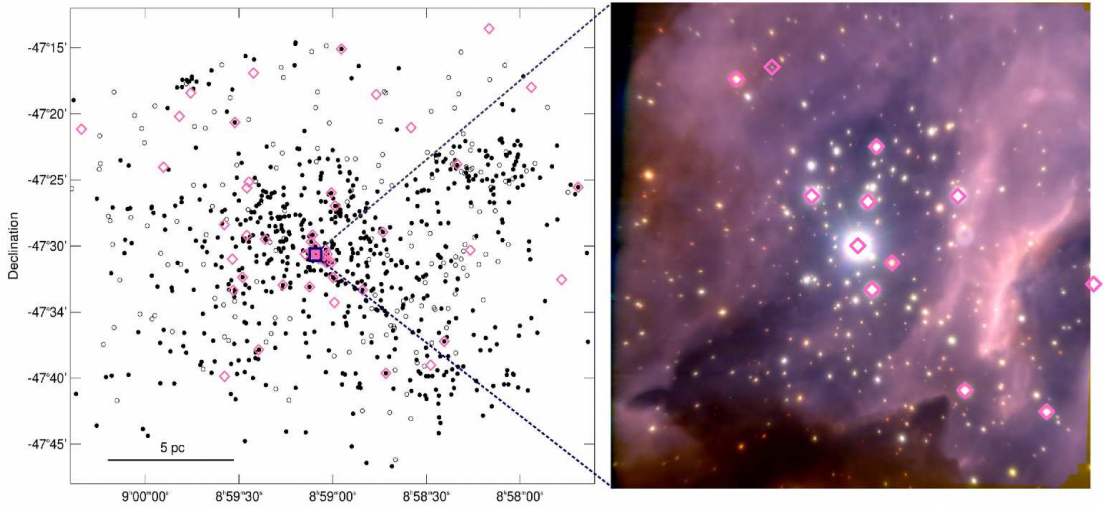}}
\caption{{\bf Left:} Large scale view of RCW\,38. Circles mark the identified YSOs (filled) and YSO candidates (open) 
from \citet{winston11}, and the diamonds mark OB star candidates \citep{wolk06, winston11}. The field covered by the NACO data 
is shown close to the center of the plot. {\bf Right:} Colour-composite image of the central part of RCW\,38, from the NACO
images in J (blue; 2003 dataset), H (green; 2013 dataset), and Ks-band (red; 2013 dataset).
The brightest star in the center of the image is the binary IRS2; the frame
size is $\sim53\times51$\,arcsec$^2$.
}
\label{fig_composite}
\end{figure*}

The left panel of Fig.~\ref{fig_composite} shows a large scale view of RCW\,38, with the circles marking the identified YSOs (filled) 
and YSO candidates (open) from \citet{winston11}, and the diamonds marking the OB-star candidates \citep{wolk06, winston11}. The small
square close to the center marks the central half parsec of the cluster studied in this work at high angular resolution. 
At least an order of magnitude denser than the nearby star forming regions, and rich in massive stars, 
RCW\,38 is an ideal environment to look for potential differences in BD formation efficiency.

This paper is structured as follows. Section~\ref{Obs&DR} contains the details of the observations and data reduction.
The data analysis, including the PSF fitting, photometric calibration, and completeness analysis, is presented in Section~\ref{analysis}. 
In Section~\ref{results}, we discuss the cluster membership, derive the mass distribution, 
test the influence of unresolved binaries on our derivation of the IMF, 
and estimate BD frequencies. In the same section we also present stellar densities in RCW\,38, along with several other nearby star forming regions.
Summary and conclusions are given in Section~\ref{summary}.

\section{Observations and Data Reduction}
\label{Obs&DR}

\begin{table*}
\caption{Summary of the observations.}
\label{tab:obs}
\begin{tabular}{lcccccccc}
\hline
Object & $\alpha$ (J2000) & $\delta$ (J2000) & date & instrument & pixel scale & FoV & filter & total exp. time\\
\hline
RCW\,38 & 08:59:05.5 & -47:30:43 & 2013-01-01, 01-12, 01-21, 01-26 & NACO & 0.054$''$ & 54$''$ & H & 6880\,s\\
RCW\,38 & 08:59:05.5 & -47:30:43 & 2012-11-21, 12-31, 2013-01-01			  & NACO & 0.054$''$ & 54$''$ & Ks & 3648\,s\\
RCW\,38 & 08:59:05.5 & -47:30:43 & 2003-02-23 & NACO & 0.054$''$ & 54$''$ & J & 448\,s\\
RCW\,38 & 08:59:05.5 & -47:30:43 & 2003-02-23 & NACO & 0.054$''$ & 54$''$ & H & 420\,s\\
RCW\,38 & 08:59:05.5 & -47:30:43 & 2003-02-23 & NACO & 0.054$''$ & 54$''$ & Ks & 480\,s\\
control field &  08:54:55.3  & -46:46:24 & 2012-10-26 & ISAAC & 0.148$''$ & 152$''$ & H & 2160\,s\\
control field &  08:54:55.3  & -46:46:24 & 2012-10-29 & ISAAC & 0.148$''$ & 152$''$ & Ks & 1890\,s\\
\hline
\end{tabular}
\end{table*}


\subsection{NAOS-CONICA/VLT}
Observations of RCW~38 were performed using the AO-assisted imager NAOS-CONICA (NACO; \citealt{lenzen03,rousset03}) at the 
ESO's Very Large Telescope (VLT), under the program number 090.C-0270. We also use the data available from the ESO archive (program ID 70.C-0400),  
previously published in \citet{derose09}. The observing setup in both programs includes the S54 camera, with the pixel scale
of 0.054$''$ and $54 \times 54$ arcsec$^2$ field of view. 
The S54 camera is accompanied by the instrument mask FLM$\_$100, which attenuates flux towards the edges of the field. The effect
is corrected for by the flat field, except in the corners of the image where the transmission falls below 5\%. 
The details of the observations, including filters and total on-source
exposure times, are listed in Table \ref{tab:obs}. The AO loop was locked on the central binary [FP74] RCW\,38 IRS 2. The binary 
separation is small enough ($0.27''$) not to be resolved by the NAOS Shack-Hartmann wavefront sensor. 

The data from 090.C-0270 were obtained in service mode, with various observing blocks executed at different nights between November 2012 and 
January 2013, and were later combined into a single deep mosaic for each band. The data from 70.C-0400 were obtained in February 2003. 
For more details about this data set please refer to \citet{derose09}. 
In the remainder of the paper, we refer to these two datasets as the ``2003 dataset'' and ``2013 dataset''.

Standard near-infrared data reduction techniques were applied using our house-brewed {\sc IDL} routines, including sky subtraction 
(obtained on a nearby dark field), flat fielding and bad-pixel correction. The 2013 dataset is affected by the so-called 50 Hz noise,
which used to sporadically appear in the data taken with the Aladdin3 detector (operational until September 2014). The noise manifests itself
as a horizontal stripe pattern, variable in intensity and time. The pattern was removed using the procedure described in \citet{hussmann12}.
The shift between the individual exposures was determined using the routine $jitter$ from the $eclipse$ package \citep{devillard97}, allowing the mosaic construction 
by a simple shift-and-add. 

The colour-composite image of the central region of RCW\,38, constructed from the J-band (blue; 2003 dataset), H-band (green; 2013 dataset), and Ks-band (red; 2013 dataset) images,   
 is shown in the right panel of the Fig.~\ref{fig_composite}.
 
\subsection{ISAAC/VLT}

To estimate the amount of contamination by field stars, we observed a control field, using the near-infrared camera ISAAC at the VLT \citep{moorwood98}. 
The Hawaii short wavelength arm of the instrument was used to obtain $H$- and $Ks$-band images, providing the field of view of 
$152\times152$ arcsec$^2$ and pixel scale of 0.148$''$ per pixel. 
A control field should be far enough from the cluster not to contain any of its stars, but also close enough to trace the same background population.  
Our control field was chosen to be located about 1 degree away from the center of the RCW$\,$38 cluster, along the galactic plane. This region
is just outside the region of significant CO emission of the Vela Molecular Ridge C \citep{yamaguchi99}, far from the dense CO clumps. 
It is also free of H$\alpha$ emitters distributed throughout the Vela Molecular Ridge \citep{pettersson08}.
The control field does not contain sufficiently bright stars that could provide good quality AO correction with NACO (the brightest star has $K_S=12.0$ and $V=14.1$). 
However, the density of stars in the control field images is only $\sim$0.04 stars per arcsec$^2$, around 4 times lower than in the cluster field.
Therefore, the seeing limited ISAAC data are suitable to estimate the field star contamination in RCW\,38.
Also, as we will show in Section~\ref{cfphot}, the control field photometry is deeper that the cluster photometry.
The details of the observations are included in Table \ref{tab:obs}. 

Standard data reduction, including sky subtraction (determined from dithered science exposures), 
flat-field and bad pixel corrections, and creation of final mosaics was performed using the ESO's ISAAC pipeline, 
and our own {\sc IDL} scripts. 

\section{Data analysis}
\label{analysis}

\subsection{PSF fitting photometry and astrometry}
\label{psf_fit}
For the NACO mosaics, photometry and astrometry were obtained using the {\sc StarFinder} PSF-fitting
algorithm \citep{diolaiti00}, particularly suitable for the analysis of AO-assisted data in crowded fields.
{\sc StarFinder}  uses an empirical PSF, extracted directly from the imaging data. This is crucial in the case of the AO data, as
the shape of the PSF can be extremely complex, especially in the presence of anisoplanatic effects. The method applied in this work is 
described in 
detail in \citet{schoedel10}, and we refer the interested reader to this work, containing a detailed justification of the method and 
particular details of the {\sc StarFinder} setup.

In our data, we observe strong anisoplanatic effects towards the edges of the detector, expected given the field of view of $\sim 55''$, and 
the isoplanatic angle in the H-band of the order $10''$ at $1''$ optical seeing. Stars at the edges of the detector are strongly elongated, and  
using a single PSF in the analysis of AO observations
with a relatively large FOV inevitably results in systematic errors in both astrometry and photometry, dependent on the radial
distance from the AO guide star \citep{schoedel10}. The effect is more strongly pronounced in the 2013 dataset. We therefore implemented the following approach:

\begin{enumerate}
\item The images were rebinned (oversampled) by a factor of 2, which helps to deal with undersampling of the NACO S54 data. 
Having well-sampled data is a necessary requirement for {\sc StarFinder}.
We then create a set of moving tiles, each with the size $512 \times 512 $ pixel$^2$ (i.e. $\sim 13.8''\times 13.8''$), and each overlapping 75\%
with the previous one. This results in $13\times13$ overlapping subfields.
The size of the subfields was chosen
 to be comparable to the size of the isoplanatic angle, while still large enough to have several stars suitable for local PSF extraction. 

\item Optimal PSF extraction requires several bright, isolated sources to be able to determine not only the central core of the 
PSF, but also the faint seeing foot further away from the center. 
In our case, the brightest stars are located closer to the center of the image.
However, the anisoplanatism will have a much more important effect on the PSF core than on the seeing foot which is caused by the light not corrected by the AO. 
Therefore, we extracted an initial PSF with well defined wings, using several very bright, isolated stars close to the cluster center. We then merged the 
individual PSF cores extracted from each sub-image, with the wings of the initial PSF. 
This step is important to avoid strong systematic variations of the zero point across the field-of-view. 

\item About 200 stars brighter than $H=16.5$, distributed across the field of view were used for the local PSF extraction. Their positions
were supplied to {\sc StarFinder}, which then performed the local PSF extraction, the merging of each local PSF core with the faint-wing PSF,
and finally the PSF source extraction for each subfield. Since we designed the fields to have a large overlap, the majority of the sources will be 
represented in several sub-images. The slightly different subset of stars used for PSF extraction between overlapping sub-images 
will affect the photometry and astrometry of each individual source. We use the values from the overlapping sub-images to calculate the 
final flux and position (average), as well as their respective uncertainties. The uncertainties have been derived 
combining in quadrature the measurement errors supplied by {\sc StarFinder} and the standard deviation of values
derived from multiple fields. The flux uncertainties given by {\sc StarFinder} take into account
Gaussian and photon noise of the images.
The corners of the image ($128 \times 128$ pix$^2$) are represented only once, therefore 
only the {\sc StafFinder} errors could be taken into account. However, they do not contain any stars due to the NACO S54 camera's field mask.


\end{enumerate}


\subsection{Photometric calibration}
\label{phot}

\begin{figure}
\centering
\includegraphics[width=0.49\textwidth]{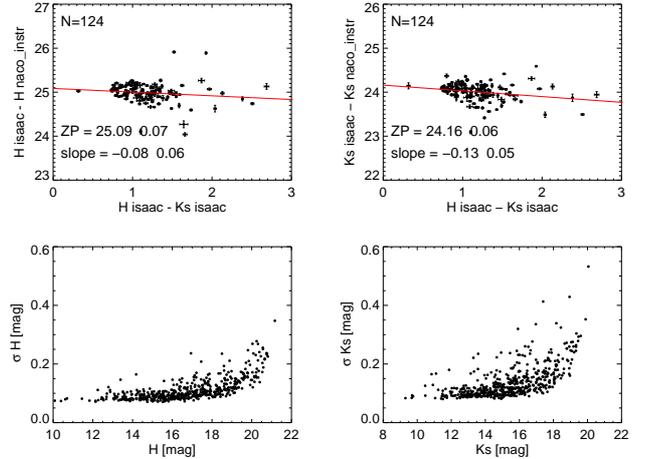}
\caption{Photometric calibration for the main 2013 dataset used in this paper. 
{\bf Upper panels:} Zero-point and colour-term fit. 
{\bf Lower panels:} 
Photometric uncertainties as a function of magnitude.  
}
\label{fig_photcal}
\end{figure}

To calibrate our photometry, we use the catalog from Ascenso et al. (in prep.), extracted from the 
larger FoV images of RCW\,38 obtained by ISAAC/VLT. The seeing-limited ISAAC data suffer more from crowding effects than
the NACO data; 
we have therefore used only matching sources without any neighbour within $0.5''$ having a luminosity contrast $\Delta H>2$\,mag.
Furthermore, we have rejected objects with uncertainties larger than 0.05 mag
in the ISAAC data, and 0.1 mag in the NACO instrumental magnitudes, and took into account only the objects brighter than $Ks$=15.
We fit the zero-points, as well as the colour terms using the following
equations:
\begin{equation}
\begin{aligned}
 J= J_{instr} + ZP_1 + c_1*colour\\
 H = H_{instr} + ZP_2 + c_2*colour\\
 Ks = Ks_{instr} + ZP_3 + c_3*colour,
\end{aligned}
\end{equation}
where $colour$=$J-Ks$ for the 2003 dataset, and $colour$=$H-Ks$ for the 2013 dataset (no $J$-band data).
In the case of 2013 dataset, only the last two equations are valid.
\begin{table*}
\centering
\caption{Parameters of the NACO photometric calibration.}
\label{tab:ZP}
\begin{tabular}{lcccccc}
dataset & ZP$_1$ & c$_1$  & ZP$_2$ & c$_2$ & ZP$_3$ & c$_3$ \\
\hline
2003 & $25.12\pm0.11$ & $0.02\pm0.03$ & $25.08\pm0.09$ & $-0.02\pm0.03$ & $24.08\pm0.09$ & $-0.02\pm0.03$\\
2013 & ... & ...  & $25.09\pm0.07$ & $-0.08\pm0.06$ & $24.16\pm0.06$ & $-0.13\pm0.05$\\
\hline 
\end{tabular}
\end{table*}
The derived ZPs and colour terms are given in Table~\ref{tab:ZP}.

The photometric uncertainties were calculated by combining the uncertainties of the zero-points,
colour-terms, and the uncertainties in {\sc StarFinder}  extraction due to PSF variation described in Section~\ref{psf_fit}.
Fig.~\ref{fig_photcal} shows the ZP and colour term fits for the 2013 dataset (upper panels), along with
the photometric uncertainties as a function of magnitude, for the $H$- and the $Ks$-band.
The 2013 combined $HKs$ catalogue contains 507 objects; 309 of these match with the $JHKs$ catalogue from 2003.
We note that in the analysis of the common sources we adopt the 2003 $J$-band photometry, combined with the 2013 photometry in $H$ and $Ks$.

\subsection{Completeness of the photometry}
\begin{figure}
\centering
\resizebox{7.5cm}{!}{\includegraphics{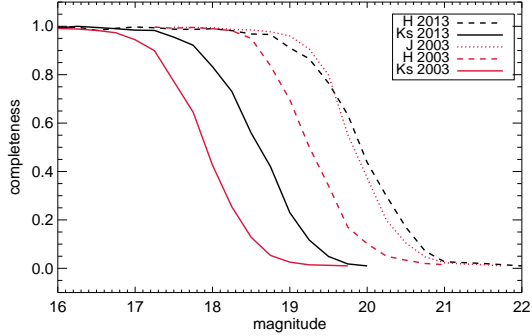}}
\caption{Completeness of the NACO photometry calculated with an artificial star experiment, for the two datasets, with 2013 shown in black, and 2003 in red. 
Different line styles represent different photometric bands: full, dashed, and dotted lines for $Ks$-, $H-$, and $J-$bands, respectively.}
\label{fig_compl}
\end{figure}

The completeness of our photometry was assessed with an artificial star test. An image containing randomly 
positioned artificial stars was created with {\sc StarFinder}'s function $image\_model$, taking into account the spatially variable PSF.
This image was added to the broad-band mosaic, and the associated photon noise was added in quadrature to the
noise image. 
The resulting image and its noise were used as the input for a routine identical to the one used to obtain the photometry with 
the spatially variable PSF (described in Section~\ref{phot}). 
We inserted only 100 stars at a time to avoid crowding, 
and repeated the same procedure 10 times at each magnitude (in steps of 0.25 mag) to improve statistics. 
The ratio between the number of recovered and inserted artificial stars versus magnitude is shown in Fig.~\ref{fig_compl}, 
for both 2003 (red lines) and 2013 datasets (black lines).
The 90$\%$ completeness limits are $H= 18.6\,$mag and $Ks=17.2\,$mag for the 2003 dataset, and $H=19.1\,$mag and $Ks=17.8\,$mag for the deeper 2013 dataset.
The completeness limits derived for the 2003 dataset are deeper than those previously reported by \citet{derose09}, likely due to our improved source
extraction technique taking into account spatial variability of the PSF.

\subsection{Control field photometry}
\label{cfphot}
The PSF photometry on the control field data was performed using {\sc StarFinder}, employing a single empirical PSF extracted from stars across the field. 
The photometric zero points were calculated from the comparison with the 2MASS catalogue \citep{2mass}. 
Stars brighter than $H\sim 14$ and $Ks\sim 13$ might suffer from non-linearity effects and were excluded. 
We use the sources  with 2MASS quality flags C or better (SNR$>5$ in all three bands), and with 
2MASS photometry errors $\leq$0.1\,mag.
Finally, we keep only the sources without a neighbour within a radius of $3''$, 
or with one that is more than 2 mag fainter. 
Due to a relatively small overlap range between the ISAAC and 2MASS catalogs for the control field, 
we only derive the zero points for the photometric calibration, ignoring any potential colour terms.

The completeness of the photometry was again assessed using an artificial star test, inserting
100 stars at a time, and repeating the procedure 10 times at each magnitude, in steps of 0.2\,mag. 
The control field photometry is 90$\%$ complete at $H=20.8$, and $Ks=20.3$.
This is 1.7 and 2.5\,mag deeper than the 2013 NACO dataset in the $H$- and $Ks$-band, respectively.

\section{Results and discussion}
\label{results}

\begin{figure*}
\centering
\resizebox{17cm}{!}{\includegraphics{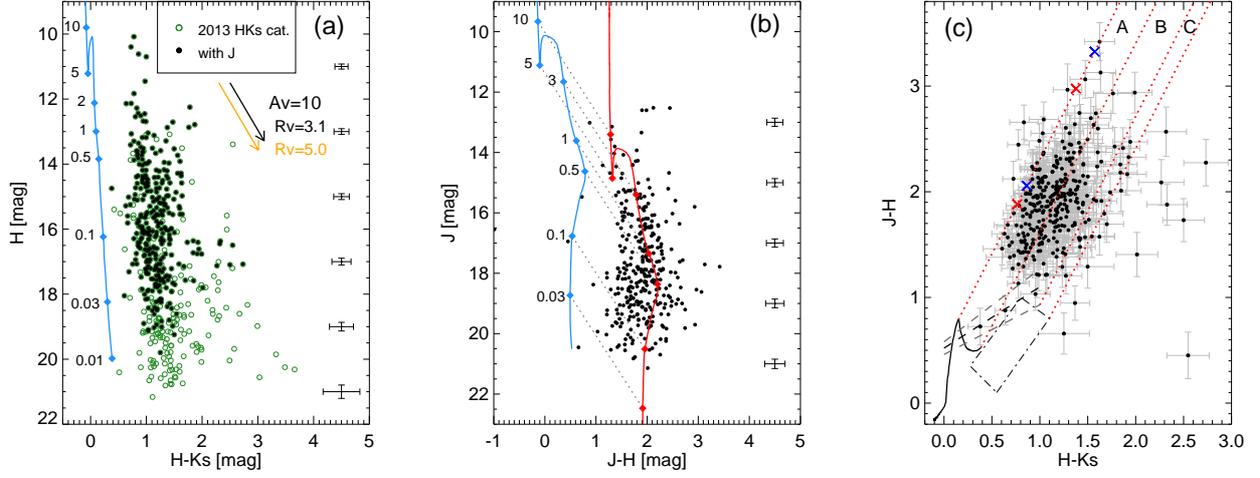}}
\caption{
{\bf(a):} $H$ vs. $H-Ks$ colour-magnitude diagram of the sources detected in RCW~38, with the green open circles marking
all the sources from the 2013 $HKs$ catalogue, and the black dots marking those that also have the $J$-band photometry available.
The blue line on the left is the combined PARSEC - BT-Settl 1 Myr isochrone. Two reddening vectors are shown, for R$_V$=3.1 (black) and R$_V$=5.0 (orange). 
{\bf(b):} $J$ vs. $J-H$ colour-magnitude diagram, with the 1 Myr isochrone at A$_V$=0 (blue) and A$_V$=13 (average extinction of the cluster for R$_V$=3.1; red).
Note that the isochrone reddened by A$_V$=11.3 and R$_V$=5.0 would be identical to the one shown here.
{\bf(c):} Colour-colour diagram. The solid black line represents the evolutionary models (PARSEC - BT-Settl), 
the dashed black and grey lines represent the locus of T-Tauri stars and the corresponding uncertainties \citep{meyer97}, whereas
 the dash-dotted rectangle represents the locus of Herbig AeBe stars \citep{hernandez05}. The dotted red lines are the reddening vectors \citep{cardelli89}.
 The crosses mark A$_V$=10 and 20 mag for R$_V$=3.1 (red) and R$_V$=5.0 (blue). 
}
\label{fig_ccd}
\end{figure*}

\begin{figure*}
\centering
\resizebox{17cm}{!}{\includegraphics{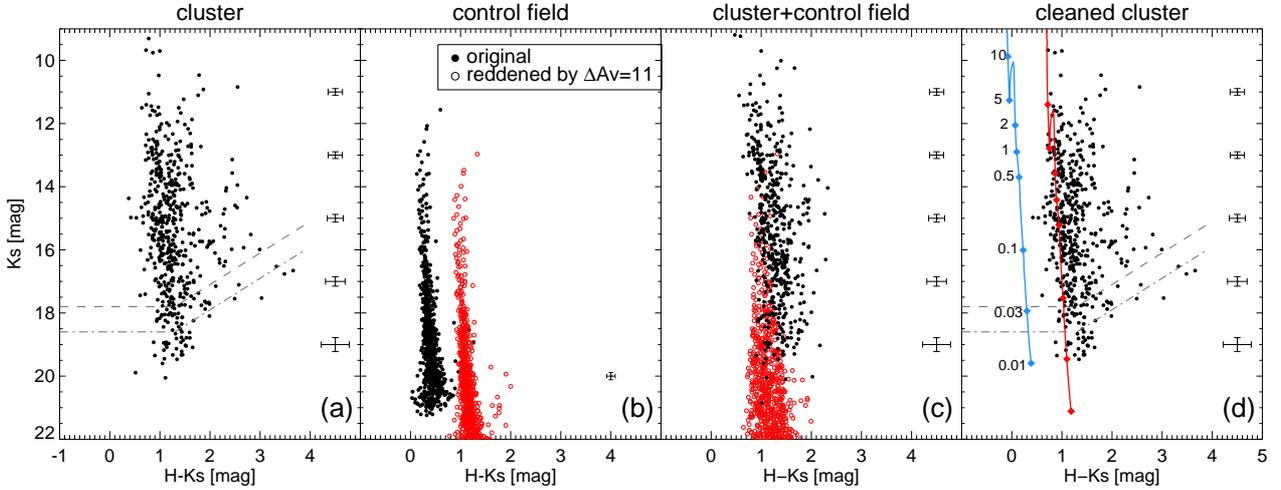}}
\caption{Colour-magnitude diagram of all the sources detected in RCW\,38 (a), and the control field (b). 
The panel (c) shows the cluster sequence (de-)reddened to the average extinction of the cluster (black), and the
control field sequence reddened to the same A$_V$, and dispersed to match the photometric errors of the cluster field.  
The panel (d) shows the cluster field after the statistical removal of the contaminants. The 1 Myr isochrones are shown at Av=0 (blue) and Av=13 
(red; average value of the extinction in the cluster).
The grey lines are marking the $90\%$ (dashed) and $50\%$ (dash-dotted) completeness limits.
}
\label{fig_cmd}
\end{figure*}


\subsection{Extinction and stellar masses}
\label{sec_mass}

Fig.~\ref{fig_ccd} shows the colour-magnitude diagrams, (CMD) and the colour-colour diagram (CCD) used to derive the extinction of the 
sources in RCW\,38. Given the young age of the cluster, many objects are expected to host circumstellar disks or envelopes, 
which in turn can introduce some unknown excess to the intrinsic near-infrared colours of cluster members. As argued by
\citet{cieza05}, classical T-Tauri stars (CTTSs) can present an excess already in the J-band, and therefore a simple de-reddening 
of the photometry to the model isochrones will overestimate the extinction, and consequently also the stellar luminosity. 
To derive the extinction towards individual stars in the cluster, we therefore develop the following procedure that takes into account a
possible intrinsic excess.

\begin{enumerate}
\item Below 1\,M$_{\sun}$, we employ the latest version of the BT-Settl models \citep{baraffe15}, while for the more massive stars we use
the PARSEC\footnote{\url{http://stev.oapd.inaf.it/cgi-bin/cmd}} stellar tracks \citep{bressan12, chen14}.
As can be appreciated from Fig.~\ref{fig_ccd}, the transition at 1\,M$_{\sun}$ between the two sets of models is smooth.
We assume a distance to the cluster of 1700 pc, an age of 1 Myr, and adopt the extinction law from 
\citet{cardelli89} with the standard value of the total-to-selective extinction ratio R$_V=3.1$, as well as R$_V=5.0$, which
might be more suitable for dust-rich environments. 
To estimate the effect that the photometric uncertainties have on our determination of extinction and mass, we apply the Monte Carlo method:
for each source we create a set of 1000 magnitudes in each band, assuming a normal distribution with a standard deviation equal
to the respective photometric uncertainty. For each of the 1000 realizations we then derive the mass and A$_V$ as described below. 
The final mass and A$_V$ for each source are calculated as an average, and their uncertainties as the standard deviation of all realizations.

\item For the sources with the full $JHKs$ photometric information (309 objects out of 507 found in the 2013 $HKs$ catalogue), 
we first check the source's position in the CCD diagram.  
In Fig.~\ref{fig_ccd}~(c), the solid black line represents evolutionary models, the dashed black and grey lines
the locus of T-Tauri stars and the corresponding uncertainties \citep{meyer97}, whereas the dash-dotted rectangle represents the locus of 
Herbig AeBe stars \citep{hernandez05}. The dotted red lines are the reddening vectors \citep{cardelli89}, encompassing the
regions where the colours are consistent with reddened evolutionary models (region A), CTTSs (region B), and Herbig AeBe stars (region C).
If the star falls in the region A of the CCD, its extinction and corresponding mass are derived by de-reddening its photometry to the
1 Myr isochrone in the $J, (J-H)$ CMD. In region B, the extinction is derived by de-reddening the colours to the CTTS locus \citep{meyer97}. 
We note that the objects in the region A could also be part of the CTTSs, however, the derived A$_V$ differences are typically smaller than
the derived uncertainties, and for simplicity we decide to only consider the evolutionary models.
The derived extinction is then used to convert the $J$-band photometry to the absolute $J$-band magnitude, which is in turn compared to the 
models to derive the mass. 
For the stars in the region B, in addition to the interstellar extinction, we also correct for an excess due to the 
circumstellar disk or envelope, chosen randomly in the interval 0 - 0.7\,mag for the $J$-band \citep{cieza05}.
In case that the derived mass is larger than 2\,M$_{\sun}$ (upper mass limit for T-Tauri stars), 
the procedure is repeated by de-reddening the colours to the Herbig AeBe locus \citep{hernandez05}. The intrinsic colour in this case is a randomly 
chosen value within the dash-dotted box shown in Fig.~\ref{fig_ccd}~(c), falling along the reddening line. The same procedure is 
performed if the objects falls within the region C. Finally, if the object falls to the left of the region A, or to the right of the region C, the extinction
and mass cannot be derived. 

\item
For the sources without the $J$-band photometry, the extinction is derived from the $H$ vs $(H-Ks)$ CMD. 
To account for a possible intrinsic infrared excess, in half of the 1000 realizations for each source, we allow for an intrinsic $(H-Ks)$ excess, as a randomly chosen value between 0.4 and 1\,mag, and the intrinsic $H$-band excess in the interval 0 -- 1.2\,mag.
The color span was chosen according to the intrinsic colours of the CTTSs (Fig.~\ref{fig_ccd}~c), and the $H$-band excess of 
CTTSs given in \citet{cieza05}.
The 50$\%$ fraction is chosen because about half of the $JHKs$ sources are found left of the region B in the CCD.  
The mass is then derived from the absolute $H-$band magnitude.

We note that the spectral types explored in \citet{cieza05} range between K0 and M4, while the least massive objects
studied here are expected to roughly be $\sim$M8 (based on the masses we derive). The CTTS sample of \citet{downes08} includes 3 BDs, which
show an $H-Ks$ excess of 0.4 - 0.6\,mag. Since we are lacking more information on the near-infrared excess at the substellar boundary, and
since the young BDs share several properties with the CTTSs (e.g. the disk fractions; \citealt{dawson13}),
we make an assumption that the 
values would not be drastically different from those of the low-mass stars.

\item
The isochrones show a degeneracy at the (pre-) main sequence transition region in the CMD, for the masses roughly between 3 and 10\,M$_{\sun}$.
For stars whose de-reddened photometry falls within this region, all the possible mass/A$_V$ combinations that match the isochrones enter the final
calculation of the star's mass and A$_V$. As a consequence, stars in this mass range have larger relative errors, and cause the 
slight ``kink'' in the IMF at masses around 5-6\,M$_{\sun}$.

\end{enumerate}  

The average extinction and the standard deviation for R$_V=3.1$ are 13.1\,mag and 7.4\,mag, respectively; 
for R$_V=5.0$ the average extinction is 11.3\,mag, with a standard deviation of 6.4\,mag. 
It is worth noting that different values of R$_V$ in the near infrared do not affect the slopes of the reddening vectors in the \citet{cardelli89} formulation of the extinction law (see Fig.~\ref{fig_ccd}), but only
their length. Therefore, the choice between the two reddening laws affects the estimate of the extinction for individual sources, but has no significant impact on the derived masses.

\subsection{Cluster membership}
To assess the membership in RCW 38 and statistically correct for the field sources, we
compare the cluster colour-magnitude diagram (CMD) with that of the control field. First, the difference in extinction
between the two lines of sight has to be taken into account. 

The fact that the $H-Ks$ colour has a relatively small scatter for sources of all spectral types
allows us to determine an average extinction along any line of sight by assuming an average
intrinsic $H-Ks$ for all the observed stars.
To determine the average extinction
in the control field, we use the method outlined in \citet{gutermuth05}. We consider the $H-Ks$ values
of the nearest $\sim$20 stars at each point on a uniform $2''$ grid. The extinction at each point of the grid
is derived as the 3$\sigma$-clipped average of the extinction values derived assuming an average intrinsic
 $H-Ks=0.25\pm 0.07$, and the extinction law of \citet{cardelli89}. The average intrinsic colour was calculated 
 using the colours of O9-M9 dwarfs from \citet{pecaut13}, and assuming the underlying mass function from \citet{chabrier05}.
Since we are not looking in the direction of the bulge ($l\sim$267$^o$), contamination by giants should be small;
$<3\%$ of the sources along the control field line of sight
are expected to be giants or subgiants, according to the Besan\c{c}on Galaxy model \citep{robin03}. 
Taking into account only the dwarf colours should therefore be a safe assumption. 
 
The average extinction in the control field derived in this way is A$_V = 2.1 \pm 1.1\,$mag for R$_V=3.1$, and A$_V = 1.9 \pm 1.0\,$mag for R$_V$=5.0. 
The difference between
the average extinctions of the cluster and the control field is therefore A$_V \simeq 11$ for R$_V=3.1$ or A$_V \simeq 9.5$ for R$_V$=5.0 (A$_{Ks}=1.3$ in both cases), the amount
by which the control field sequence has to be shifted in the CMD before comparison. 
The panel (a) of Fig.~\ref{fig_cmd} shows all the observed objects in the direction of the cluster field. The grey lines mark the $90\%$ (dash-dotted), 
and $50\%$ (dashed) completeness limits. The control field CMD is shown in the panel (b), with the light grey symbols
showing the original photometry derived for this field, and the red ones the same sequence reddened by A$_V = 11$. 

The CMD is subdivided into grid cells with a step size of 0.5\,mag in both axes. The number of field stars
within each cell is normalized to the ratio of the on-sky areas between the cluster and the control field, 
and corrected for the completeness fractions of the cluster field photometry (the reddened control field photometry is 100$\%$
complete below the cluster faint end). 
The resulting number of objects corresponding to the field object population
is then randomly subtracted from corresponding cells of the cluster CMD.
However, before the correction, there are two effects that have to be taken into account. The first is the fact
that the cluster suffers a significant differential reddening across the field, which means that only the population 
with extinction close to the average one would be taken into account for the correction. To avoid this, we should (de-)redden the cluster population 
to the average cluster extinction (A$_V$=13 for R$_V=3.1$). To take into account the uncertainties in A$_V$, the photometry of each star is de-reddened
to a random value within $\pm 1\sigma$ from the average cluster extinction, where $\sigma$ is the star's A$_V$ uncertainty (black dots in panel (c) of Fig.~\ref{fig_cmd}). 
The sources for which we could not derive the extinction in Section~\ref{sec_mass} have been removed prior to this step (3 out of 507).
The second effect is that the control field sequence has significantly lower uncertainties, 
especially at the faint end, which again will cause some cells to be under- or over-corrected. To fix this, we artificially
disperse the control field sequence to match the photometric uncertainties of the cluster field (red dots in panel (c) of Fig.~\ref{fig_cmd}).   
The resulting
field-subtracted cluster population is shown in the righ-most panel of Fig.~\ref{fig_cmd}, along with the 1 Myr isochrones 
at Av=0 (blue) and Av=13 (red).
As evident from Fig.~\ref{fig_cmd}, only a small fraction of the sources gets removed from the cluster sequence by
this method ($\sim 5\%$ of the detected objects). As previously discussed by \citet{derose09}, the cluster seems to be located
in front of a dense molecular cloud whose high extinction screens out background sources. The cleaned cluster 
sequence contains 476 candidate members.

\begin{figure}
\centering
\resizebox{8cm}{!}{\includegraphics{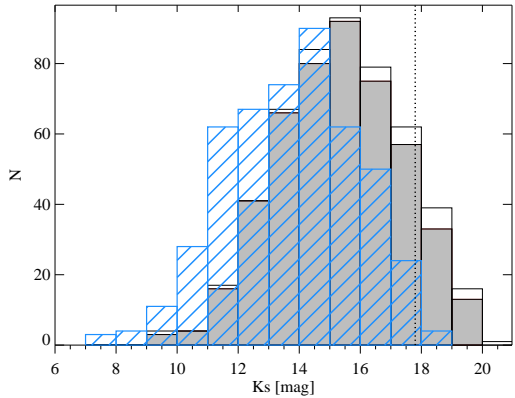}}
\caption{$Ks$-band luminosity function. The white histogram shows all the sources from the $HKs$ catalogue, while
the grey one shows the probable member sample determined through comparison of the cluster CMD with that of the control field. 
The blue hatched histogram shows the extinction-corrected version of the grey one. The dotted line marks the 90$\%$ completeness limit.
}
\label{fig_klf}
\end{figure}

Fig. ~\ref{fig_klf} contains the $Ks$-band brightness distribution, with the statistically cleaned cluster population shown in grey. The hatched 
blue histogram contains the same data as the grey one, corrected for extinction. Fitting a Gaussian to the extinction-corrected histogram,
we get the peak located at $Ks=13.9$,
which at the distance of 1.7 kpc corresponds to an absolute magnitude of $M_{Ks}$=2.75, or $\sim 0.4\,$M$_{\sun}$.
This is similar to what is seen in other young clusters, e.g., Trumpler\,14 ($0.4-0.5\,$M$_{\sun}$ at a distance 2.3 - 2.8 kpc and 
$\langle $A$_V \rangle =3$; \citealt{ascenso07, rochau11}), or Trapezium ($\sim0.5$\,M$_{\sun}$ for d$=0.4$\,kpc and 
$\langle $A$_V \rangle =9.2$; \citealt{muench02}). 
It is also in agreement with the simulated K-band luminosity function for clusters of non-accreting PMS stars,
which peak at M$_K=2.5-3$ for the ages 0.7 - 1 Myr  \citep{zinnecker91}.

\subsection{Initial Mass Function}
\label{sec:imf}

\begin{figure*}
\centering
\resizebox{17cm}{!}{\includegraphics{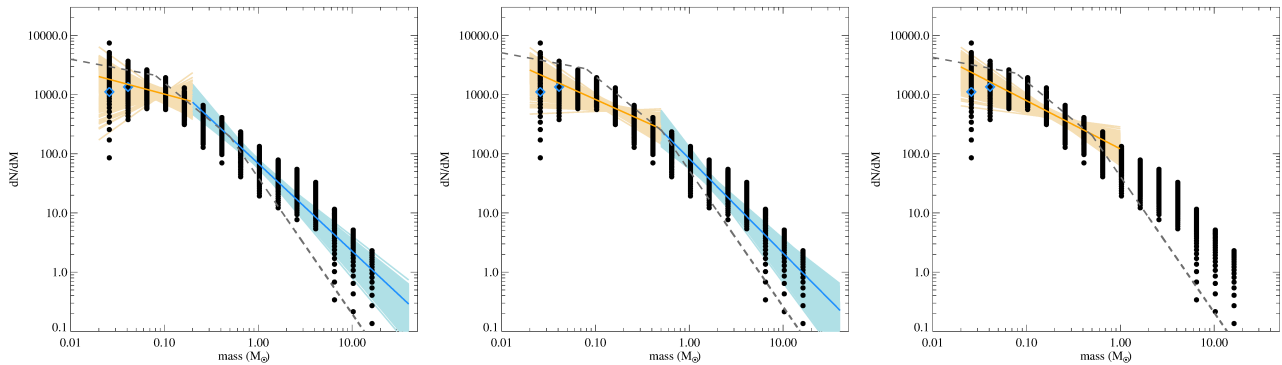}}
\caption{Initial mass function of the central core of RCW\,38, represented in the equal-size bins of 0.2 dex in mass. We plot the points (black) and the 
corresponding fits from $10^4$ realizations of the performed Monte Carlo simulation (light blue and orange shaded area). 
The dark blue/orange lines show the final IMF, with the slopes calculated as a weighted average of the slopes obtained in the simulation.
The IMF is represented as a two-segment power-law,
with different break masses shown in the three panels. The blue diamonds mark the median value of all the realizations without correcting for the completeness,
for the bins where it has effect. The grey dashed line is the Kroupa segmented power-law mass function normalized at 0.5\,M$_{\sun}$.
}
\label{fig_imf}
\end{figure*}

\begin{table}
\centering
\caption{Parametrization of the Initial Mass Function in RCW\,38 in the power-law form, for a selection of different 
mass break points shown in Fig.~\ref{fig_imf}.}
\label{tab:imf}
\begin{tabular}{ccc}
mass range [M$_{\sun}$] & $\alpha$  & $\alpha$  \\
\hline
						& R$_V$=3.1 & R$_V$=5.0 \\	
\hline
\hline
		0.2 - 20 		& $1.48 \pm 0.08$  & $1.49 \pm 0.09$  \\
		0.02 - 0.2	 	& $0.42 \pm 0.18$ & $0.45 \pm 0.18$ \\	
\hline
		0.5 - 20 	 	& $1.60 \pm 0.13$ & $1.60 \pm 0.13$ \\
        0.02 - 0.5 		& $0.71 \pm 0.11$& $0.72 \pm 0.11$ \\	  
\hline
		0.02 - 1		& $0.81 \pm 0.08$ & $0.83 \pm 0.09$ \\ 
\hline 
\end{tabular}
\end{table}

To plot the IMF for RCW\,38 we take the statistically cleaned probable member masses as derived in previous sections 
and run a Monte Carlo simulation where each mass is moved by a random offset within its respective $\pm1 \sigma$ uncertainty.
This is performed 100 times, and for each of the 100 realizations we do 100 bootstraps, i.e., random sampling with replacement.
In other words, starting from a sample with N members, in each bootstrap we draw a new sample of N members, allowing some members of the initial sample to be 
drawn multiple times.
 This results in
10$^4$ mass distributions, which are used to derive the final IMF and its uncertainties. 
In Fig.~\ref{fig_imf} we show the IMF and the corresponding fits in the power-law form $dN/dM\propto M^{-\alpha}$. 
A Poisson uncertainty is assigned to each fitted data point, but is not shown in the plot for clarity.
The lowest masses derived for the objects in the sample are $\sim\,$0.02\,M$_{\sun}$, setting the lower limit
for the IMF.
The data in the lowest-mass bins  
have been corrected for incompleteness, according to the average $Ks$-band magnitude of the sources belonging to it, and the curve shown
in Fig.~\ref{fig_compl}. The blue diamonds mark the median of the uncorrected values for the two lowest mass bins, where the correction is significant. 
We fit a two-segment power-law using the least-square method, with the break masses at 0.2 (left panel in Fig.~\ref{fig_imf}) and 0.5 M$_{\sun}$ (middle panel).
For a comparison, we overplot the Kroupa segmented power-law IMF, normalized at 0.5\,M$_{\sun}$ (dashed grey line), with the slopes $\alpha=2.3$ for M$>0.5$M$_{\sun}$, $\alpha=1.3$ for 
$0.08<M<0.5\,$M$_{\sun}$, and $\alpha=0.3$ for M$<0.08$M$_{\sun}$.
In the right-most panel we show the fit only below 1\,M$_{\sun}$, to be able to compare it with the IMF fits
in the nearby young clusters from the SONYC survey. The summary of the resulting slopes is given in Table~\ref{tab:imf}. 
The results are largely insensitive to R$_V$.

On the high-mass side, the slope of the power-law in the central $\sim0.45\times0.45$\,pc$^2$ of RCW\,38 is 
found to be $\alpha=1.48 \pm 0.08$, and $1.59 \pm 0.13$, for the masses above 0.2\,M$_{\sun}$, and 0.5\,M$_{\sun}$, respectively.
This is lower than the Salpeter slope ($\alpha$=2.35). A similar slope has been 
reported in the central regions of other massive clusters. 
In the central region ($r \leq 0.2\,$pc) of the Arches cluster, \citet{habibi13} find the slope $\alpha=1.5\pm0.35$, while
\citet{andersen16} report a slope $\alpha = 1.32 \pm 0.16$ in the inner 0.8\,pc radius of Westerlund 1.  
In the inner 0.2\,pc radius of NGC\,3603, the observed IMF slope is $\alpha\simeq1.3$ \citep{harayama08, stolte06}.  
In $\sigma\,$Ori, \citet{penaramirez12} measure a slope $\alpha = 1.7 \pm 0.2$ for masses above 0.35\,M$_{\sun}$, while
in Collinder~69 \citet{bayo11} report $\alpha\sim1.8$ for the mass range 0.65 - 25\,M$_{\sun}$.
In Arches, Westerlund 1, and NGC\,3603, the above mentioned works all report a steepening of the observed slope
at larger radii (to values from 1.8 to Salpeter), which is interpreted as evidence for mass segregation. However, 
great care should be taken when interpreting these results in distant, rich clusters, because incompleteness due to crowding 
can produce a similar effect of radially changing slope, even 
in a completely un-segregated cluster \citep{ascenso09}. As the latter work shows, 
a rigorous study of completeness (which is not always found in the literature) is a pre-requisite when analyzing the radial properties 
in crowded regions. 

In the case of RCW\,38, although the crowding does not pose a real problem in the presence of the AO observations, the present work does not allow us
to judge whether the cluster is mass segregated, simply because
the mass function of the wider RCW\,38 is yet to be derived.
The combined X-ray ($Chandra$) and mid-infrared ($Spitzer$) observations reveal a presence of 4 sub-clusters \citep{winston11}. 
The central sub-cluster surrounding the source IRS\,2 shows a higher concentration of OB candidates than any of the other
three sub-clusters, however, an even larger number of OB candidates are found distributed throughout the extended cluster region. 
\citet{kuhn14} describe the RCW\,38 morphology as a ``core-halo'' structure, meaning that overall the cluster structure
appears unimodal with little substructure around a main cluster, but with an excess of stars near the center. In 
\citet{kuhn15b} the authors estimate the relaxation time for the core, coinciding roughly with the regions studied here, to be $\sim 5\,$Myr. 
For the wider cluster area ($\sim 4.5\,$pc radius), the relaxation time is much longer, $>200\,$Myr. Given its very young age ($\lesssim1\,$Myr), RCW\,38 is therefore likely 
not a dynamically relaxed system.

\begin{table*}
\caption{Power-law slope $\alpha$ of the low-mass IMF in various star forming regions ($dN/dM\propto M^{-\alpha}$).}
\label{tab:slopes}
\begin{tabular}{llccc}
\hline
Region & Mass range [M$_{\sun}$] & $\alpha$ & Isochrones				& Reference\\
\hline
NGC\,1333 		& 0.005 -- 0.60		& $0.6\pm0.1$	& NextGen + Dusty				& \citet{scholz12a} \\
				& 0.030 -- 1.00		& $0.9 - 1.0$	& BT-Settl, Dusty, NextGen		& \citet{scholz13} \\
Chamaeleon-I 	& 0.005 -- 1.00		& $0.78\pm0.08$ & BT-Settl 						& \citet{muzic15} \\
Lupus~3 		& 0.020 -- 1.00 	& $0.79\pm0.13$	& BT-Settl 					& \citet{muzic15} \\
Collinder\,69 	& 0.010 -- 0.65		& $0.18 - 0.38$	& Siess + Cond					& \citet{bayo11} \\
$\sigma$~Ori	& 0.006 -- 0.35		& $0.55\pm0.20$	& Dusty + Cond					& \citet{penaramirez12} \\
				& 0.006 -- 0.35		& $0.68\pm0.23$	& Dusty + Cond + Siess			& \citet{penaramirez12} \\
				& 0.006 -- 0.11     & $0.6\pm0.2$	& Dusty + Cond + NextGen		& \citet{caballero07}\\
				& 0.012 -- 0.10		& $0.7 \pm 0.3$ & Dusty + Cond + NextGen    	& \citet{bejar11}\\
IC\,348 		& 0.012 -- 0.075	& $0.7\pm0.4$	& Dusty							& \citet{ado13}  \\
				& 0.030 -- 1.00		& $0.7 - 0.8$	& BT-Settl, Dusty, NextGen		& \citet{scholz13}  \\
$\rho$\,Oph 	& 0.004 -- 0.075	& $0.7\pm0.3$	& Dusty							& \citet{ado12} \\
ONC 			& 0.030 -- 0.30		& $-0.12 \pm 0.90 $	& NextGen					& \citet{dario12} \\
				& 0.020 -- 0.17		& $-1.41 \pm 0.25 $	& DM97						& \citet{dario12} \\
				& 0.006 -- 0.10		& $0.31\pm0.11$ &  	NextGen + Dusty				& \citet{lucas05} \\
				& 0.025 -- 0.12     & $\simeq 0.3 $  & 	DM97						& \citet{muench02} \\
				& 0.012	-- 0.15			& $0.3 - 0.6$	& NextGen + Dusty				& \citet{weights09} \\				
Upper Scorpius 	& 0.009 -- 0.2		& $0.45\pm0.11$	&  	BT-Settl					& \citet{lodieu13}$^{*}$\\
\hline
\end{tabular}
\footnotetext{}{BT-Settl:~\citet{allard11}; Siess: \citet{siess00}; COND: \citet{baraffe03}; DUSTY: 
\citet{chabrier00}; NextGen: \citet{baraffe98}}; DM97: \citet{dantona97}\\
\footnotetext{}{$^{*}$ using the 5 Myr isochrone}
\end{table*}

On the low-mass side, we can compare the results in RCW~38 given in Table~$\ref{tab:imf}$ with the results of numerous
surveys dealing with the low-mass content in nearby star forming regions and young clusters, including our 
SONYC survey. To facilitate the comparison, we compile a summary of various results that extend into the BD regime, provided in Table~\ref{tab:slopes}.
We include mass ranges in the quoted power-law fits, as well as the employed isochrones. 
The slope $\alpha$ is in agreement between most of the listed star forming regions and clusters.
 Our results in RCW\,38 
are also similar to what is found in other regions, i.e. {\it our data leave no evidence for an effect that a combination of high stellar densities
and OB star presence might have on efficiency of BD/very-low mass star production.}

The only cluster with comparable age and stellar density to RCW\,38, and studied in the substellar regime so far is the ONC. 
Despite being one of the best studied young clusters in general, its low-mass IMF remains controversial. 
There have been several proposals of a bimodal mass function, 
with a second peak, or a hint of it,  in the substellar regime \citep{muench02, slesnick04, lucas05, drass16}, 
where only the work of \citet{slesnick04} includes spectroscopic confirmation of the substellar candidates.
\citet{dario12} report an IMF that declines steeply in the substellar regime, with a negative $\alpha$ slope, in contrast to
$\alpha\sim\,$0.3 - 0.6 reported by  \citet{muench02, weights09}. 
Possible explanations for these discrepancies include membership issues, issues with understanding survey completeness, different
 methodologies (optical vs infrared photometry, (non-)existence of the spectroscopic follow-up, adoption of
different isochrones and extinction laws), as well as the distance uncertainties (values between 400 and 480 pc assumed in the mentioned works).
Moreover, \citet{andersen11} report a gradient in the number ratio of stars (0.08 - 1 M$_{\sun}$) to BDs (0.03 - 0.08 M$_{\sun}$),
suggesting that the low-mass content of the cluster is mass segregated. Significantly different areas covered in some of the above mentioned surveys 
 might therefore influence the derived mass functions to some extent.
Finally, there might be an additional complication in the ONC, in the form of a slightly older (4-5 Myrs)
population located in front of the Orion A cloud \citep{alves&bouy12, bouy14}. This population overlaps significantly with the ONC, and can make up for more than
10-20\% of the ONC population, and even 30-60\% excluding the Trapezium cluster \citep{alves&bouy12}.
Given the youth of the foreground population, and the similarity of its radial velocity to that of the Trapezium population,
the membership analysis is far from trivial, and probably affects the IMF determination.

As evident from Table~\ref{tab:slopes}, the slope $\alpha$ of the mass function in the power-law form $dN/dM\propto M^{-\alpha}$ 
is larger than zero, in contrast to the analysis by \citet{andersen08}. Based on the number ratios of stars to BDs in seven different 
star forming regions, they conclude that the substellar IMF turns over with a slope $\alpha < 0$. The reason for this inconsistency might be 
an overestimate of star-to-BD ratios for some of the clusters, due to incompleteness and/or small number statistics. 
For example, for IC~348, \citet{andersen08} quote a ratio of $\sim 8$, while the newer surveys yield the values 3-4 \citep{scholz13}.

Finally, it is worth mentioning that most mass functions that we compare here are system mass functions, with no correction for unresolved
binaries, justifying a direct comparison with the RCW\,38 IMF. On the other hand, the differences between the system and a single star IMF
are definitely worth exploring, which is what we do in Section~\ref{binaries}.

\subsection{Star/BD ratio}

A useful quantity for the comparison of IMF measurements from different surveys
is the number ratio of stars to BDs. To estimate the ratio, we use masses 
derived in Section~\ref{sec_mass}, and apply the same method as we did in deriving the IMF, generating 10$^4$ mass distributions. 
For the stellar-substellar mass boundary, we take the value at the solar metallicity, 0.075\,M$_{\sun}$. The low mass limit on the 
BD side is fixed to 0.03\,M$_{\sun}$, and the 
upper limit for the stellar mass range is chosen to be 1\,M$_{\sun}$. These limits have been chosen to match those
most commonly used in the literature.
We derive the star to BD number ratio in RCW\,38 of $2.1 \pm 0.6$. The large relative uncertainty of the ratio is mostly 
due to the incompleteness levels on the low-mass side, and the photometric uncertainties at the star-BD border. 
The star to BD ratio in RCW\,38 is similar to that of NGC\,1333 ($1.9-2.4$; \citealt{scholz13}), somewhat lower than that of IC\,348 ($2.9 - 4$; \citealt{scholz13}),
and overlaps with the lower values of the range calculated for Cha-I and Lupus~3 ($2.5 - 6.0$; \citealt{muzic15}). However, as pointed out in \citet{muzic15}, the numbers in Cha-I and Lupus~3 
might in reality be on the lower side of the quoted span, as suggested by the analysis based on the completeness levels of the spectroscopic follow-up.

\subsection{Comparison of stellar densities}
\label{densities}
In Section~\ref{sec:imf} we compare the low-mass IMF in RCW\,38 to the mass functions found in other (less dense) nearby
star forming regions. In this section we try to quantify the differences in the stellar densities between several regions. 
We calculate the stellar surface and volume  densities in RCW\,38, and three well studied nearby clusters,
Chamaeleon-I, NGC 1333, and the ONC, representing the loose, intermediate and dense stellar environments, respectively. The area for the calculation has a radius of 0.2\,pc,
which is dictated by the field-of-view of the RCW\,38 dataset. For the volume densities, we assume a uniform, spherical distribution with the radius of 0.2\,pc.
Given the relatively small radius used in the calculation, the values
 derived here can be considered core densities for each cluster. 

We take into account only stars more massive than 0.1\,M$_{\sun}$, to avoid errors due to incompleteness.
 For RCW\,38, we use the masses calculated in 
this work. For Cha-I, we consider the masses calculated in \citet{muzic15}, which are based on the census combined from several works \citep{luhman04a,luhman07,luhman&muench08, daemgen13, muzic15}.
For NGC\,1333 we take the census from \citet{luhman16}, and exclude all the objects with the spectral type M7 or later. An M7 object should have the Teff$\sim$2900\,K \citep{muzic14}, 
equivalent to 0.1\,M$_{\sun}$ at 1 Myr, according to the BT-Settl models. The NIR sources without spectral classification brighter than $Ks$=12.2 are kept, since they
are expected to be more massive than 0.1\,M$_{\sun}$ already at A$_V=0$. There are several YSOs detected only in the MIR, which might be embedded protostars. Excluding, or 
including these sources gives us a range of densities for NGC\,1333.  
For the ONC, we take the masses from \citet{dario12}, using the \citet{baraffe98} models.

For RCW\,38, the center is set to the brightest object, the IRS2 binary. At the distance of 1700 pc, we obtain the stellar surface density $\Sigma\sim$2500\,pc$^{-2}$, and 
stellar volume density $\rho\sim$16500\,pc$^{-3}$. 
This is about 6 times lower than the average surface density within the same region found by (\citealt{kuhn15}; $\Sigma$ = 15 800 stars\,pc$^{-2}$)
The discrepancy might be caused by the correction of the surface density maps with the spatially dependent detection fraction in \citet{kuhn15}, which are 
derived from the X-ray observation, and typically drop at the positions of cluster cores.
Cha-I has no obvious center looking at the overall region, and consists of two regions with stellar clustering. We take 
the centers of the two parts as defined in \citet{king12a}, and get $\Sigma\sim$95\,pc$^{-2}$ and $\rho\sim$640\,pc$^{-3}$ for the northern component, and 
$\Sigma\sim$70\,pc$^{-2}$ and $\rho\sim$480\,pc$^{-3}$ for the southern component, assuming the distance of 160 pc. Similar to Cha-I, NGC\,1333 does not
have a well defined center: we consider two regions with the highest concentration of members: the first one is center on the bright F-type star
located at 03:29:10.38, +31:21:59.2, while the other is centered at 03:28:57,+31:16:50.9. 
We assume a distance of 300 pc, and obtain $\Sigma\sim$180 - 220\,pc$^{-2}$ and $\rho\sim$1200 - 1450\,pc$^{-3}$. For the ONC, we center the field to the center of 
the Trapezium cluster, adopt the distance of 400 pc, and obtain $\Sigma\sim$1000\,pc$^{-2}$ and $\rho\sim$6800\,pc$^{-3}$.
We note that the values for ONC are in agreement with those calculated by \citet{king12a}, while those in Cha-I are slightly higher due to the 
updated census used here. 

The central part of RCW\,38 is therefore more than two times denser than the ONC, an order of magnitude denser than NGC\,1333, and about 25 times denser than Cha-I.

\subsection{Effect of the unresolved binaries on the IMF}
\label{binaries}
A significant fraction of stars host one of more companions, which, in the case they are 
unresolved, might affect our determination of the IMF. In this section we
 model this effect and the uncertainty it has on the IMF, assuming certain
binary frequencies and mass ratios (described below), as well as excluding the possibly resolved wide systems. 

\subsubsection{Binary frequencies}
\citet{king12a,king12b} collected binary statistics for seven young star forming regions, with the separation range of $62-620\,$AU
and primary mass range of $0.1 - 3.0\,$M$_{\sun}$ 
being directly comparable between five of them. In this separation range, the lowest 
density region, Taurus, has a binary fraction of $\sim 21\%$, while 
the other 4 regions, with densities greater than a few 100 stars\,pc$^{-3}$ (including the ONC), show a lower fraction of $\sim\,10\%$.
Since the density of RCW\,38 is comparable to that of the ONC, and given that this is one of the most thoroughly studied nearby regions, 
we will use the available ONC statistics for our assumptions on the binary properties of the RCW\,38 population. 

\citet{koehler06} studied the multiplicity properties for 275 objects in the ONC, in the separation range 60-500 AU. 
In the core of the cluster, they find the binary fraction of $2.3\pm3.0 \%$ in the mass range 0.1\,M$_{\sun}\leq$M$\leq2\,$M$_{\sun}$,
and $21.1\pm10.2 \%$ for M$\geq 2\,$M$_{\sun}$. The most massive star in their sample has a mass estimate of 16\,M$_{\sun}$.
Based on a sample of $\sim780$ ONC members, \citet{reipurth07} derive the binary fraction of $8.8\pm1.1 \%$, in  
the separation range 67.5-675 AU. Although they give no information of the mass range that was studied, almost all
identified binaries with existing spectral type estimate have a spectral types K-M.
In a recent work based on the $HST$ data, \citet{kounkel16b} analyze 324 protostars and PMS stars in Orion (not only the ONC), 
and find a companion fraction\footnote{Given that only 2 out of 57 identified multiple systems are higher order systems (triples), 
the binary and companion fractions should be similar in this case.} of $10.3\pm1.0\%$ in the separation range 100-1000 AU. Interestingly, 
they also find a higher companion fraction in denser regions ($\sim 13.5$\% vs $\sim 8$\% for stellar surface density $\Sigma > 45\,$pc$^{-2}$ and $< 45\,$pc$^{-2}$, respectively),  
in contrast to the results of \citet{koehler06}.

Considering binaries closer than the visual ones studied in the works mentioned so far, there are several studies dealing
with spectroscopic binaries in Orion. \citet{morrell91} find the binary frequency for periods shorter than
100 days ($\lesssim1\,$AU for a BV star) among the main-sequence members of the association is 32$\%$. 
The majority
of the studied stars are massive, predominantly BV type. On the other hand, \citet{abt91} find the binary frequency among the 
26 brightest ONC members (spectral types O6-A1) to be 15$\%$. 
\citet{kounkel16a}, find the multiplicity frequency of $5.8\pm1.1\%$, 
for a sample of (typically) K-type stars, i.e. on average lower mass than the previous two samples, and up to 10\,AU separation. 
From the high resolution optical interferometry of a small sample of 9 brightest stars in the ONC (A0 - O5.5), \citet{grellmann13}
find 6 multiple systems, and conclude that the massive stars in the ONC have a companion fraction 5 times higher than the low mass stars. 
The separations probed in this work are $\sim 1-80\,$AU, filling the gap between the close binaries probed by the radial velocities and AO surveys.

We note that the works dealing with visual binaries normally assume that the the projected separation represents the true semi-major axis, although this is
statistically not exactly correct \citep{kuiper35}. Also, they typically do not correct for the fact that at inclined orbits, a companion  
will appear for a certain fraction of its period at projected separations shorter than the spatial resolution of the observations. 
However, as we will show in Section~\ref{widesystems}, for our resolution limit of 270 AU, this correction should be around $10-15\%$. For the ONC
surveys, which probe much shorter separations, this number will be even smaller, therefore it is probably safe to ignore it.

In summary, stars more massive than $\sim 2$\,M$_{\sun}$ have binary frequency estimates varying between 15 and 60\%. 
On the low-mass side, both close ($<10\,$AU) and wide ($>60$\,AU) companions in the ONC appear with the frequency of $5-10\%$. In the separation range
of 10-60\,AU, however, we have no available data for the ONC. 
For the main sequence stars and field BDs, the orbital period distribution can be approximated with a log-normal distribution that peaks at $\sim$5 AU for stars with the spectral type M or later, and at $\sim$45 AU for FGK stars \citep{duchene&kraus13}. 
We might therefore be missing a substantial number of binaries in the separation range not covered by the current surveys.
For this reason, we also test the binary frequency of $20\%$.

\subsubsection{Mass ratios}
Given the uncertainties in mass determinations in young clusters, mass ratio distribution $q$ might be a more complex parameter to probe than the
binary frequency. In a review paper on stellar multiplicity, \citet{duchene&kraus13} show that the distribution 
of mass ratios (in form $dN/dq \propto q^\gamma$,) down to $q\sim$0.1 for all masses above 0.3\,M$_{\sun}$ is essentially flat ($\gamma \sim 0$), whereas 
below this limit it becomes increasingly skewed towards high-q systems ($\gamma \sim 3$). 


\subsubsection{Resolved wide systems}
\label{widesystems}
The resolution of the NACO data is on average $\sim$160 mas (calculated as $2\times$FWHM of the variable PSFs across the field), which is 
equivalent to $\sim$270\,AU at the distance of 1.7 kpc. 
To determine how many wide multiple systems we might have detected in our images, we searched for candidate companions within
$0.5''$ ($\approx 850\,$AU) of each source in the NACO images. We identify 32 potential binary and triple systems (35 if we count the 3 triples as 2 binaries each), 
with the separations between 
290 and 820 AU, and the median of 590 AU. Of the 32 systems, 20 have estimated primary masses below 2\,M$_{\sun}$. First, we have to explore the possibility that these are line of sight pairs rather than genuine binaries.
To do so, we determine the stellar surface density $\Sigma$ in an area with a radius of $10''$ ($\approx17000$ AU) around each candidate primary, 
and calculate the probability P of finding an unrelated star within a distance $\theta$ from each primary by $P=1-e^{-\pi \theta^2 \Sigma}$
\citep{correia06,reipurth07}, where $\theta=0.5''$. We find a non-negligible probability of chance alignment for all our
candidates, between 7\% and 26\%, with an average of $\sim 20\%$. This means that $\sim$26 of the detected pairs might be physical binaries. 
Since in this analysis we deal with projected separations, rather than true binary orbit sizes, we might also ask how many
of the visual binaries we miss due to inclination effects, where a companion  
 appears, for a certain fraction of its period, at projected separations shorter than the spatial resolution of the observations. 
To estimate how often this happens, we setup a simple geometric simulation in which a set of circular orbits
with semi-major axis 270 - 1100 AU is run through a set of inclinations. The upper limit comes from our chosen upper separation limit of 850 AU, multiplied by  
the average ratio between the semi-major axis and its projected separation for an ensemble of binaries, which is 1.3 \citep{kuiper35}. 
We find that the probability of the two components being at projected distances below 270 AU is $\sim$10\%. Therefore, $\sim18$ low-mass and $\sim11$ high-mass 
visual binaries should be excluded from
our simulation.
We can repeat the same analysis for the range of up to 620\,AU (useful for the comparison with \citealt{king12b}, see below): we find 
20 multiple systems, of which $\sim 11\%$ are chance alignments, and $\sim$15\% are missing due to the projection effects.

Comparing the separation distribution for 7 star forming regions, \citet{king12b} find that
regions with higher densities exhibit a similar proportion of wide (300-620\,AU) relative to close
(62-300\,AU) binaries, which might be unexpected from the preferential destruction of wider pairs.
Binaries with separation 300-620\,AU comprise $10 - 40\,\%$ of the total binary populations in the 7
regions studied in \citet{king12b}, with the ONC value being $\sim 25\%$.
In RCW\,38, we detect 20 candidate multiple systems, of which 10 have primary masses $>2$\,M$_{\sun}$. Subtracting the 
probable chance alignments, we get the multiplicity frequency in the 300-620\,AU range of 10/147 and 10/360, for the high-
and low-mass bins, respectively. Assuming that these comprise $25\%$ of all the 60-620\,AU binaries, would 
imply total binary frequencies in this mass range of $\sim 27\%$ for masses $>2$\,M$_{\sun}$, and $\sim 11\%$ for masses $<2$\,M$_{\sun}$.
This is roughly in agreement with the binary frequencies in the ONC listed in the previous section.

\begin{table*}
\centering
\caption{Results of the single-star IMF simulation, for a set of different binary frequencies. The system IMF is shown for comparison.}
\label{tab:bin}
\begin{tabular}{llccccccc}
 Case      & 		\multicolumn{2}{c}{Binary fraction}                                     	& \multicolumn{5}{c}{$\alpha$}\\
    \hline
       & 	&	 & 0.02-0.2\,M$_{\sun}$&  0.2-20\,M$_{\sun}$ & 0.02-0.5\,M$_{\sun}$ & 0.5-20\,M$_{\sun}$  & 0.02-1\,M$_{\sun}$ \\

\hline
\hline
system IMF & & & $0.42\pm0.18$ & $1.48\pm$0.08 & 0.71$\pm$0.11 & 1.60$\pm$0.13 & 0.81$\pm$0.08 \\ 
\noalign{\smallskip}
  I	   & \begin{tabular}{@{}c@{}}M$<2\,$\,M$_{\sun}$ \\ M$\geq2\,$M$_{\sun}$ \end{tabular} & \begin{tabular}{@{}c@{}}0.05 \\ 0.15 \end{tabular}  & $0.42\pm 0.19$& $1.50\pm0.08$ &    $0.69 \pm 0.11$            &  $1.62 \pm 0.13$              & $0.80 \pm 0.09$  \\
 \noalign{\smallskip } 
  II   & \begin{tabular}{@{}c@{}}M$<2$\,M$_{\sun}$ \\ M$\geq2$\,M$_{\sun}$ \end{tabular} & \begin{tabular}{@{}c@{}}0.10 \\ 0.15 \end{tabular}  & $0.45\pm0.17$ & $1.50\pm0.08$ &    $0.73\pm 0.11$            &  $1.62 \pm 0.12$              & $0.83 \pm 0.09$  \\
 \noalign{\smallskip}  
  III  & \begin{tabular}{@{}c@{}}M$<2$\,M$_{\sun}$ \\ M$\geq2$\,M$_{\sun}$ \end{tabular} & \begin{tabular}{@{}c@{}}0.10 \\ 0.30 \end{tabular}  &$0.46 \pm 0.17$ & $1.57\pm0.09$&    $0.73 \pm 0.10$            &  $1.76 \pm 0.13$              & $0.83 \pm 0.08$  \\
  \noalign{\smallskip}
  IV  & \begin{tabular}{@{}c@{}}M$<2$\,M$_{\sun}$ \\ M$\geq2$\,M$_{\sun}$ \end{tabular} & \begin{tabular}{@{}c@{}}0.10 \\ 0.60 \end{tabular}  &  $0.46\pm0.18$ & $1.58\pm0.09$  &    $0.75 \pm 0.11$            &  $1.78 \pm 0.14$              & $0.82 \pm 0.09$  \\
 \noalign{\smallskip}
  V  & \begin{tabular}{@{}c@{}}M$<2$\,M$_{\sun}$ \\ M$\geq2$\,M$_{\sun}$ \end{tabular} & \begin{tabular}{@{}c@{}}0.20 \\ 0.60 \end{tabular}   &  $0.51\pm0.17$ & $1.59\pm0.09$ &  $0.77 \pm 0.11$            &  $1.77 \pm 0.13$              & $0.85 \pm 0.09$  \\
 
\hline

\end{tabular}
\end{table*}

\subsubsection{IMF simulation}
We test the following cases in our simulation: for the low-mass stars ($<2$\,M$_{\sun}$), 
we test binary fractions of 5, 10$\%$, and 20$\%$, whereas for the stars with masses $>2$\,M$_{\sun}$, 
we test binary fractions of 15\%, 30\%, and 60\%.
From our candidate member catalogue of masses, we randomly select a fraction that will be split into binaries. For the selected sources with mass $<0.3$\,M$_{\sun}$,
we generate a mass ratio distribution according to $dN/dq \propto q^\gamma$, where $\gamma=3$, which prefers equal-mass binaries. For masses $>0.3$\,M$_{\sun}$,
we assume a flat mass ratio distribution ($\gamma = 0$). Mass ratios are assumed to have values in the range 0.1 - 1. 
The selected sources are then split into binaries according to the corresponding mass ratio, with a requirement that the sum of the expected fluxes for the two
components matches the flux of the observed unresolved system. 
For simplicity, we keep the same A$_V$ as previously derived. 
The next step is to exclude the detected wide binary systems, by removing 11 randomly selected binaries in the high mass bin, and 18 in the low-mass one, as derived in the previous section.    

With this new set of masses, we repeat a similar Monte Carlo simulation as the one described in Section~\ref{sec:imf}.
Starting from the same catalogue as before, each mass is moved by a random offset within its respective $\pm1 \sigma$ uncertainty.
This is performed 100 times, and in each step a new subset of binaries is generated as described above. For each of these
100 realizations, we do 100 bootstraps. 
The results for the same mass intervals as those given for the system IMF (Table~\ref{tab:imf}), are shown in Table~\ref{tab:bin}.
By comparing the numbers derived for the system IMF with the results of this simulation, we see that 
unresolved binaries cause the slope on both high-mass and low-mass side to appear flatter than if the binaries were resolved. However, the
effect is not very pronounced, with the power-law indices of the single-star IMF and the system IMF agreeing within the errors.

\section{Summary and Conclusions}
\label{summary}

In this work we have presented new, deep AO observations of the central $\sim 0.5\,$pc
of the young, embedded cluster RCW\,38, taken with NACO/VLT. The depth of the data allowed us to
extend the analysis of the IMF to the substellar regime for the first time in this cluster, and to our knowledge
in any other massive young cluster at the distance above 1 kpc. The analysis is facilitated by the fact that the
cluster sits in front of a dense molecular cloud, which blocks the light from most of the background sources. The comparison 
with the control field shows that $\sim96\%$ of the observed objects are expected to be cluster members.

The main goal of this work was to study the low-mass part of the IMF, and compare it with the well studied
mass distributions in nearby star forming regions. According to various BD formation theories, stellar density
is expected to affect the production of very-low mass objects (high densities favor higher production rate of BDs), as 
well as the presence of massive OB stars capable of stripping the material around nearby pre-stellar cores until leaving
an object not massive enough to form another star. In addition to the theoretical expectations, we discussed several observational hints 
for environmental differences in the nearby star forming regions from the literature, which, however, require further investigation. To that end, we choose
to study a cluster that is several orders of magnitude denser than any of the nearby star forming regions (except for the ONC), 
and also, unlike them, rich in massive stars.

Based on the sample of 476 candidate cluster members, we derive the IMF in RCW\,38 between 0.2 and 20\,M$_{\sun}$, and fit a two-segment power-law.
For the masses in the range 0.5 -- 20\,M$_{\sun}$, we find the slope $\alpha = 1.60 \pm 0.13$, shallower than the Salpeter slope ($\alpha=2.35$), but 
in agreement with several other works, mainly in the centers of the mass-segregated cores of Milky Way's starburst clusters. 
At the low-mass side, we find $\alpha = 0.71 \pm 0.11$ for masses between 0.02 and 0.5\,M$_{\sun}$,
or $\alpha = 0.81 \pm 0.08$ for masses between 0.02 and 1\,M$_{\sun}$. This is in agreement  with the values found in other young star forming regions, 
{\it leaving no evidence for environmental differences in the efficiency of the production of BDs and very-low mass stars
possibly caused by high stellar densities or a presence of numerous massive stars.}

We investigate the effects that the unresolved binaries might have on the IMF slope, assuming the binary
properties similar to the ONC, and excluding the resolved wide multiple systems. Our Monte-Carlo simulation 
reveals that the the unresolved binaries affect the resulting IMF slope, by making the system IMF shallower.
However, the effect is not very pronounced, with the power-law indices of the single-star IMF and the system IMF agreeing within the errors.

With the inclusion of RCW38, star forming regions covering a wide range
of initial conditions have now been investigated in the substellar domain.
In all regions studied so far, the star/BD ratio is between 2 and $\sim5$, 
i.e. for each 10 low-mass stars between 2 and 5 BDs are expected 
to be formed. This is also consistent with estimates of the star/BD ratio
in the field ($\sim5$; \citealt{bihain&scholz16}). The sum of these results clearly shows that brown dwarf
formation is a universal process and accompanies star formation in
diverse star forming environments across the Galaxy.

By combining previous measurements from the literature, \citet{licquia15} present improved estimates of several global properties of the Milky Way, 
including its current star formation rate (SFR) of $1.65\pm 0.19$\,M$_{\sun}$\,yr$^{-1}$.
To do so, they assume an underlying Kroupa IMF \citep{kroupa01}, which has an average stellar mass of $\sim 0.4\,$M$_{\sun}$ \citep{marks&kroupa12}.
The Milky Way therefore forms on average $4\pm0.5$ stars each year, which yields the present brown dwarf formation rate of 0.7 - 2.3 objects per year.
Assuming that the brown dwarf formation rate remained the same through the age of the Galaxy 
yields the number of brown dwarfs in the Milky Way between 10 and 30 billions. 
However, as suggested by several works, the star formation rate of the Galaxy seems to have been larger in the past.
Simulations from the latest version of the Besan\c{c}on Galaxy model \citep{czekaj14}, in comparison
to the Tycho-2 catalogue data, result in the SFR that decreases with time, no matter which IMF is assumed.
In their case, the best fit to the data is provided by the exponentially decreasing star formation rate $SFR(t)\propto exp(-\gamma t)$; where $\gamma=0.12$, $t$ is the time, and $t_{max}$=12.54\,Gyrs \citep{aumer&binney09}.
\citet{snaith15} developed a chemical evolution model to reconstruct the star formation history of the Milky Way's disk. Their SFR shows 
 two distinct phases that correspond to the formation of the
thick and the thin discs. The thick-disc formation lasts 4--5 Gyr,
during which the SFR reaches 10--15\,M$_{\sun}$\,yr$^{-1}$. After this first phase, 
star formation stalls for about 1 Gyr and then resumes
for the thin-disc phase at a level of 2\,M$_{\sun}$\,yr$^{-1}$
for the remaining 7 Gyr.

If we assume the brown dwarf formation rate behaving in the same way as the SFR, and again assuming the star/BD ratio of 2 -- 5, we arrive at the total number of 
BDs in the Galaxy of 24 -- 60 billions for the exponentially decaying SFR, and 30--100 billions for the two-phase disk formation model. 
These numbers should be considered as a lower limit, because the star/BD ratio used
in the calculation was derived only for the BDs more massive than 0.03\,M$_{\sun}$.

\vspace{0.5cm} 

KM acknowledges funding by the Joint Committee of ESO/Government of Chile, and by the Science and Technology Foundation of Portugal (FCT), grant No. IF/00194/2015.
Part of the research leading to these results has received funding from the European Research Council 
under the European Union's Seventh Framework Programme (FP7/2007-2013) / ERC grant agreement No. [614922].
RJ acknowledges support from NSERC grants. JA acknowledges funding by the Science and Technology Foundation of Portugal (FCT), grant No. SFRH/BPD/101562/2014.












\bsp	
\label{lastpage}
\end{document}